\documentclass[twocolumn,showpacs,preprintnumbers,amsmath,amssymb,superscriptaddress]{revtex4}

\usepackage{graphicx}
\usepackage{bm}

\def\mib#1{\hbox{\boldmath $#1$}}

\def\eq#1{Eq.\,(\ref{#1})}

\def\bfsigma{\mib{\sigma}}

\def\bn{\mib{n}}
\def\bp{\mib{p}}
\def\bq{\mib{q}}
\def\br{\mib{r}}

\def\bS{\mib{S}}

\def\CT{{\cal T}}
\def\CO{{\cal O}}

\begin{document}

\preprint{APS/123-QED}

\title{Spin-orbit splitting
of \mib{\hbox{}^9_\Lambda \hbox{Be}} excited states
studied with the \mib{SU_6} quark-model baryon-baryon
interactions}

\author{Y. Fujiwara}
\affiliation{Department of Physics, Kyoto University, 
Kyoto 606-8502, Japan}%
 \email{fujiwara@ruby.scphys.kyoto-u.ac.jp}

\author{M. Kohno}%
\affiliation{Physics Division, Kyushu Dental College,
Kitakyushu 803-8580, Japan}

\author{K. Miyagawa}%
\affiliation{Department of Applied Physics,
Okayama Science University, Okayama 700-0005, Japan}

\author{Y. Suzuki}%
\affiliation{Department of Physics, Niigata University,
Niigata 950-2181, Japan}

\date{\today}

\begin{abstract}
The previous Faddeev calculation of the
two-alpha plus $\Lambda$ system
for $\hbox{}^9_\Lambda \hbox{Be}$ is extended to incorporate
the spin-orbit components of the $SU_6$ quark-model baryon-baryon
interactions. We employ the Born kernel
of the quark-model $\Lambda N$ $LS$ interaction,
and generate the spin-orbit component
of the $\Lambda \alpha$ potential
by the $\alpha$-cluster folding. The Faddeev calculation
in the $jj$-coupling scheme implies that the direct use of the
quark-model Born kernel for the $\Lambda N$ $LS$ component is
not good enough to reproduce the small experimental
value $\Delta E^{\rm exp}_{\ell s}=43 \pm 5$ keV for
the $5/2^+$-$3/2^+$ splitting.
This procedure predicts three to five times larger values
in the model FSS and fss2. The spin-orbit contribution from
the effective meson-exchange potentials in fss2 is argued to be
unfavorable to the small $\ell s$ splitting, through the analysis of
the Scheerbaum factors for the single-particle spin-orbit potentials
calculated in the $G$-matrix formalism.
\end{abstract}

\maketitle


In the previous paper \cite{BE9L}, we have carried out
Faddeev calculations of the two-alpha
plus $\Lambda$ ($\alpha \alpha \Lambda$) system,
starting from the quark-model
hyperon-nucleon ($YN$) interaction.
The input was the Minnesota three-range $NN$ force 
for the $\alpha \alpha$ resonating-group method (RGM) kernel
and a two-range Gaussian $\Lambda N$ potential (SB potential)
which is generated from the phase-shift behavior
of the quark-model $YN$ interaction, fss2 \cite{fss2,B8B8},
by using an inversion method based on supersymmetric quantum
mechanics \cite{SB97}.
The Faddeev calculations using these central forces
were carried out in the $LS$-coupling scheme, and reproduced
the ground-state energy and excitation energies
of the $\hbox{}^9_\Lambda \hbox{Be}$ hypernucleus
within 100 $\sim 200$ keV accuracy.

Another important experimental information
from $\hbox{}^9_\Lambda \hbox{Be}$ is the small spin-orbit
splitting of the $5/2^+$ and $3/2^+$ excited states,
$\Delta E^{\rm exp}_{\ell s}=43 \pm 5$ keV \cite{AK02,TA03},
measured from the recent Hyperball $\gamma$-ray spectroscopy. 
It is widely known that the single particle (s.p.) spin-orbit
interaction of the $\Lambda$ hyperon seems to be extremely small,
especially in light $\Lambda$-hypernuclei.
In the non-relativistic models of the $YN$ interaction,
this is a consequence of the strong cancellation of
the ordinary $LS$ component and the antisymmetric $LS$
component ($LS^{(-)}$ force), the latter of which is
a characteristic feature of baryon-baryon interactions
between non-identical baryons. For example, the $SU_6$ quark-model
baryon-baryon interaction, FSS \cite{FSS},
yields a strong $LS^{(-)}$ component \cite{NA93},
which is about one half of the ordinary $LS$ component,
with the opposite sign.
We performed the $G$-matrix calculation in symmetric nuclear
matter, using this quark-model baryon-baryon
interaction \cite{GMAT}, and calculated the so-called
Scheerbaum factor, $S_B$, which indicates
the strength of the s.p. spin-orbit interaction \cite{SPLS}. 
The ratio of $S_B$ to the nucleon
strength $S_N \sim -40~\hbox{MeV}\cdot\hbox{fm}^5$ is $S_\Lambda/S_N
\sim 1/5$ and $S_\Sigma/S_N \sim 1/2$ in the Born
approximation. The $G$-matrix calculation of the model FSS
modifies $S_\Lambda$ to $S_\Lambda/S_N \sim 1/12$.
The significant reduction of $S_\Lambda$ in the $G$-matrix
calculation of FSS is traced back to the enhancement of the
antisymmetric $LS$ component in the diagonal $\Lambda N$ channel,
owing to the $P$-wave $\Lambda N$-$\Sigma N$ coupling.

Hiyama {\em et al.} \cite{HI00} calculated the $\Lambda N$ spin-orbit
splitting in $\hbox{}^9_\Lambda \hbox{Be}$ and
$\hbox{}^{13}_\Lambda \hbox{C}$ in their cluster model,
by using simple approximations of the Nijmegen
one-boson-exchange $\Lambda N$ interactions.
They employed several two-range Gaussian $LS$ potentials
for the $\Lambda N$ interaction, which simulate
the $LS$ and $LS^{(-)}$ parts of the $G$-matrix
interactions derived from Nijmegen model-D (ND), model-F (NF),
and NSC97a-f interactions.
For example, they obtained $\Delta E_{\ell s}=0.16$ MeV
for NSC97f. When the $LS^{(-)}$ force is switched off,
they obtained 0.23 MeV. Since these values are too large to compare
with the experiment, they adjusted the strength
of the $LS^{(-)}$ potentials,
guided by the relative strength of the
quark-model $LS^{(-)}$ force. Such a procedure, however,
does not prove the adequacy of the quark-model
spin-orbit interaction for the experimental data.

The purpose of this brief report is to show that,
if we carry out more serious calculations
starting from the the quark-model baryon-baryon interactions,
the situation is not so simple as stated in Ref. \cite{HI00}. 
Here we concentrate only on the spin-orbit interaction, and
use the quark-model exchange kernel directly, following our basic
idea in other applications of our $SU_6$ quark-model
baryon-baryon interactions \cite{triton,hypt,HE6LL}.
The $\Lambda \alpha$ spin-orbit interaction is generated
from the Born kernel of the $\Lambda N$ $LS$ quark-model
interaction, and the Faddeev equation is solved
in the $jj$-coupling scheme,
by using the central plus spin-orbit $\Lambda \alpha$ interactions.
We find that our model FSS yields
spin-orbit splittings of almost 2/3 of the Nijmegen NSC97f result.
We find a large difference between FSS and fss2 for the effect
of the short-range correlations, especially, in the way
of the $P$-wave $\Lambda N$-$\Sigma N$ coupling.

We assume that the $\Lambda N$ $LS$ interaction
is given by the Born kernel of the $\Lambda N$ quark-model
interaction \cite{SPLS}:
\begin{eqnarray}
v^{LS}_{\Lambda N}(\bq_f, \bq_i) & = & \sum_\Omega \sum_\CT
\left[\,(X^\Omega_\CT)^{ud}f^{LS}_\CT(\theta) \right.
\nonumber \\
& & \left. +(X^\Omega_\CT)^s\,f^{LS}_\CT(\pi-\theta)\,\right]
\CO^\Omega\ , 
\label{ls1}
\end{eqnarray}
where $\Omega=LS$, $LS^{(-)}$, and $LS^{(-)}\sigma$, specify
three different types of spin-orbit operators,
$\CO^{LS}=i \bn \cdot \bS$, $\CO^{LS^{(-)}}=i \bn \cdot \bS^{(-)}$,
and $\CO^{LS^{(-)}\sigma}=i \bn \cdot \bS^{(-)}\,P_\sigma$,
and $\CT$ stands for various interaction types originating from
the quark antisymmetrization. Here we use the standard
notation, $\bn=[\bq_i \times \bq_f]$, $\bS=(\bfsigma_\Lambda
+\bfsigma_N)/2$, $\bS^{(-)}=(\bfsigma_\Lambda-\bfsigma_N)/2$,
and $P_\sigma=(1+\bfsigma_\Lambda \cdot \bfsigma_N)/2$, etc.
The up-down and strange spin-flavor factors,
$\left(X^\Omega_\CT\right)^{ud}$ and $\left(X^\Omega_\CT\right)^s$,
in the $\Lambda N$ channel, and the direct and exchange spatial functions,
$f^{LS}_\CT(\theta)$ and $f^{LS}_\CT(\pi-\theta)$ with
$\cos \theta=(\widehat{\bq}_f \cdot \widehat{\bq}_i)$, are explicitly given
in Refs. \cite{NA93} and \cite{SPLS}.
If we take the matrix element of \eq{ls1} with
respect to the spin-flavor functions
of the $\Lambda \alpha$ system,
the nucleon spin operator part disappears due to
the spin saturated property of the $\alpha$ cluster
and we obtain the spin-flavor part
as $X^d_\CT \bS_\Lambda$ and $X^e_\CT  \bS_\Lambda$ with
$\bS_\Lambda=\bfsigma_\Lambda/2$,
$X^d_\CT=4[(X^{LS}_\CT)^{ud}
+(X^{LS^{(-)}}_\CT)^{ud}]$, and
$X^e_\CT=4[(X^{LS}_\CT)^s+(X^{LS^{(-)}\sigma}_\CT)^s]$.
We therefore only need to calculate the spatial integrals
of $f^{LS}_\CT(\theta)\,i\bn$ and $f^{LS}_\CT(\pi-\theta)\,i\bn$.
For this calculation, we can use a convenient formula Eq.\ (B.8)
given in Appendix B of Ref.\ \cite{BE9L}.
The calculation is carried out analytically,
since it only involves Gaussian integration.
We finally obtain 
\begin{eqnarray}
& & V^{LS}_{\Lambda \alpha}(\bq_f, \bq_i)
=\sum_\CT \left[\,X^d_\CT~V^{LS\,d}_\CT (\bq_f, \bq_i) \right.
\nonumber \\
& & \left. + X^e_\CT~V^{LS\,e}_\CT (\bq_f, \bq_i)\,\right]
~i \bn \cdot \bS_\Lambda\ .
\label{ls2}
\end{eqnarray}

We calculate the spin-flavor factors and spatial integrals
for each of the interaction types,
$\CT=D_-$, $D_+$ and $S$ ($S^\prime$).
From our previous paper \cite{NA93},
we find the spin-flavor factors given in Table \ref{table1}.
Note that the most important knock-on term of the $D_-$ type
turns out to be zero in the $\Lambda \alpha$ direct potential,
because of the exact cancellation between
the $LS$ and $LS^{(-)}$ factors in the up-down sector.
As the result, the main contribution to the $\Lambda \alpha$
spin-orbit potential in the present formalism
comes from the strangeness exchange $D_-$ term,
which is non-local and involves very strong momentum dependence.
If the quark mass ratio, $\lambda=(m_s/m_{ud})$,
goes to infinity, all of these spin-flavor factors vanish,
which is a well known property of the spin-flavor $SU_6$ wave
function of the $\Lambda$ particle.
Only the strange quark of $\Lambda$ contributes
to the spin-related quantities
like the magnetic moment, since the up-down di-quark
is coupled in the spin-isospin zero for $\Lambda$.
The explicit expressions of the spatial integrals,
$V^{LS\,d}_\CT (\bq_f, \bq_i)$ and $V^{LS\,e}_\CT (\bq_f, \bq_i)$,
will be given elsewhere, since they are rather lengthy.
The partial wave components of \eq{ls2} are calculated
from the formula in Appendix C of Ref.\ \cite{LSRGM},
by using the Gauss-Legendre 20-point
quadrature formula.
Since the model fss2 contains the $LS$ components
from the effective meson-exchange potentials (EMEP's),
we should also include these contributions
to the $\Lambda \alpha$ spin-orbit interaction.
A detailed derivation of the EMEP Born kernel
for the $\Lambda \alpha$ system is deferred to a separate paper.

\begin{table}[t]
\caption{The spin-flavor factors of the $\Lambda \alpha$
potential for the quark-model $LS$ exchange kernel.
The parameter, $\lambda=(m_s/m_{ud})$, is the strange to up-down
quark mass ratio.}
\label{table1}
\begin{center}
\renewcommand{\arraystretch}{1.4}
\setlength{\tabcolsep}{5mm}
\begin{tabular}{ccc}
\hline
$\CT$ & $X^d_\CT$ & $X^e_\CT$ \\
\hline
$D_-$ & 0 & ${4 \over 9\lambda}\left(2+{1 \over \lambda}\right)$ \\
$D_+$ & $-{2 \over 9\lambda}\left(2+{1 \over \lambda}\right)$
& 0 \\
$S$,~$S^\prime$
& $-{1 \over 9\lambda}\left(2-{1 \over \lambda}\right)$
& ${2 \over 9\lambda}\left(2-{1 \over \lambda}\right)$ \\
\hline
\end{tabular}
\end{center}
\end{table}

For the Faddeev calculation, we use the same conditions as
used in Ref.\ \cite{BE9L}, except for the exchange mixture
parameter $u$ of the SB $\Lambda N$ potential.
We here use a repulsive $\Lambda N$ odd interaction
with $u=0.82$, in order to reproduce the ground-state energy
of $\hbox{}^9_\Lambda \hbox{Be}$.
This is because the $5/2^+$ - $3/2^+$ $\ell s$ splitting is
rather sensitive to the energy positions of these states,
measured from the $\hbox{}^5_\Lambda \hbox{He}+\alpha$ threshold.
We also use Nijmegen-type $\Lambda N$ potentials
from Ref.\ \cite{HI97}.
The $\alpha \alpha$ RGM kernel is generated
from the three-range Minnesota force with $u=0.94687$.
The harmonic oscillator width parameter
of the $\alpha$-cluster is assumed
to be $\nu=0.257~\hbox{fm}^{-2}$. The partial waves
up to $\lambda_{\rm Max}={\ell_1}_{\rm Max}=6$ are
included both in the $\alpha \alpha$
and $\Lambda \alpha$ channels.
The momentum  discretization points are selected
by $n_1$-$n_2$-$n_3$=10-10-5 with the midpoints
$p,~q=1$, 3, and 6 $\hbox{fm}^{-1}$.
The Coulomb force is incorporated in the cut-off Coulomb
prescription with $R_C=10$ fm.

Table \ref{table2} shows the results of
Faddeev calculations in the $jj$-coupling scheme.
First we find that the ground-state energies do not change much
from the $LS$-coupling calculation, which implies the
dominant $S$-wave coupling of the $\Lambda$ hyperon.
The final values for the $\ell s$ splitting
of the $5/2^+$ - $3/2^+$ excited states
are $\Delta E_{\ell s}=137$ keV for FSS and 198 keV for fss2,
when the SB force with $u=0.82$ is used
for the $\Lambda N$ central force.
If we compare these results with the experimental value,
$43 \pm 5$ keV, we find that our quark-model predictions
are three to five times too large.
If we use the $G$-matrix simulated NSC97f $LS$ potential
in Ref.\ \cite{HI00}, we obtain 209 keV for the same
SB force with $u=0.82$. The difference from 0.16 MeV
in Ref.\ \cite{HI00} is due to the model dependence 
to the $\alpha \alpha$ and $\Lambda \alpha$ interactions.
We find that our FSS prediction
for $\Delta E_{\ell s}$ is less than 2/3 of the NSC97f prediction,
while fss2 gives almost the same result as NSC97f. 
If we switch off the EMEP contribution
in the fss2 calculation, we find $\Delta E^{\rm exp}_{\ell s}=86$ keV.
This results from the very small $LS^{(-)}$ component generated
from the EMEP of fss2.  

\begin{table}[t]
\caption{The ground-state energy $E_{\rm gr}(1/2^+)$,
the $5/2^+$, $3/2^+$ excitation energies $E_{\rm x}(5/2^+)$,
$E_{\rm x}(3/2^+)$, and the spin-orbit
splitting, $\Delta E_{\ell s}=E_{\rm x}(3/2^+)-E_{\rm x}(5/2^+)$,
calculated by solving the Faddeev equations
for the $\alpha \alpha \Lambda$ system
in the $jj$ coupling scheme.
The exchange mixture parameter of the SB $\Lambda N$ force
is assumed to be $u=0.82$. The $\Lambda \alpha$ spin-orbit force
is generated from the Born kernel of the FSS and
fss2 $\Lambda N$ $LS$ interactions.
For the fss2 $LS$ interaction, the $LS$ component
from the EMEP is also included.
}
\label{table2}
\begin{center}
\renewcommand{\arraystretch}{1.2}
\setlength{\tabcolsep}{2mm}
\begin{tabular}{cccccc}
\hline
$V^{LS}_{\Lambda N}$ & $V^{\rm C}_{\Lambda N}$
& $E_{\rm gr}(1/2^+)$
& $E_{\rm x}(5/2^+)$ & $E_{\rm x}(3/2^+)$
& $\Delta E_{\ell s}$ \\
& & (MeV) & (MeV) & (MeV) & (keV) \\
\hline
     & SB & $-6.623$ & 2.854 & 2.991 & 137 \\ 
     & NS & $-6.744$ & 2.857 & 2.997 & 139 \\
FSS  & ND & $-7.485$ & 2.872 & 3.024 & 152 \\
     & NF & $-6.908$ & 2.877 & 3.002 & 125 \\
     & JA & $-6.678$ & 2.866 & 2.991 & 124 \\
     & JB & $-6.476$ & 2.858 & 2.980 & 122 \\
\hline
     & SB & $-6.623$ & 2.828 & 3.026 & 198 \\ 
     & NS & $-6.745$ & 2.831 & 3.033 & 202 \\
fss2 & ND & $-7.487$ & 2.844 & 3.064 & 220 \\
     & NF & $-6.908$ & 2.853 & 3.035 & 182 \\
     & JA & $-6.679$ & 2.843 & 3.024 & 181 \\
     & JB & $-6.477$ & 2.834 & 3.012 & 178 \\
\hline
\multicolumn{2}{c}{Exp't~\protect\cite{TA03}}
 & $-6.62(4)$ & 3.024(3) & 3.067(3)
 & $43(5)$ \\
\hline
\end{tabular}
\end{center}
\end{table}

\begin{figure}[t]
\begin{minipage}[t]{85mm}
\includegraphics[angle=-90,width=82mm]{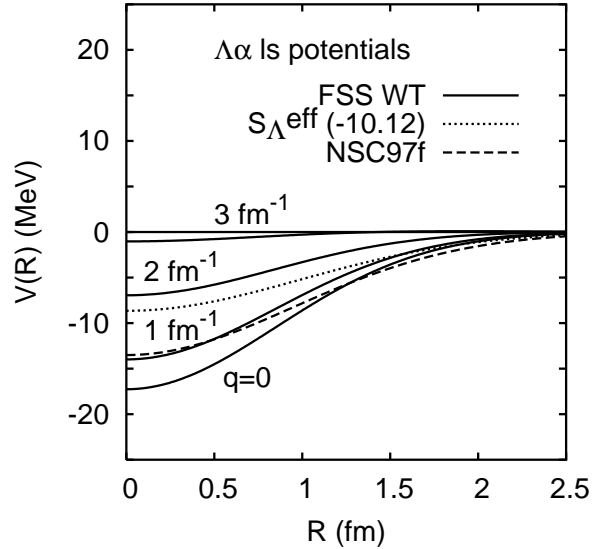}
\caption{
Comparison of $\Lambda \alpha$ spin-orbit potentials
predicted by the Wigner transform of FSS
with $q=0,~1,~2$, $3~\hbox{fm}^{-1}$, and
$\widehat{\br} \cdot \widehat{\bq}=0$ (solid curves),
the Scheerbaum potential with $S^{\rm eff}_\Lambda
=-10.12~\hbox{MeV}\cdot \hbox{fm}^5$ (dotted curve),
and the $G$-matrix simulated NSC97f-type
potential \protect\cite{HI00} (dashed curve).
}
\label{fig1}
\end{minipage}
\end{figure}

As an alternative prescription to correlate
the $\hbox{}^9_\Lambda \hbox{Be}$ $\ell s$ splitting and the
basic baryon-baryon interaction, we examine
the Scheerbaum's s.p. spin-orbit potential (Scheerbaum potential)
for the $(0s)^4$ $\alpha$-cluster in the Scheerbaum formalism.
For the Scheerbaum factor,
$S_\Lambda=-8.3~\hbox{MeV}\cdot \hbox{fm}^5$, calculated
in Ref.\ \cite{SPLS} for FSS in the simplest Born approximation,
we obtain $\Delta E_{\ell s}=121$ keV for the SB force
with $u=0.82$. If we compare this with the value 137 keV in
Table \ref{table2} for FSS, we find that
the Scheerbaum potential seems
to be reliable even in our quark-model nonlocal kernel.
However, this agreement is fortuitous, since the center-of-mass
correction to the $\ell s$ operator in the $\Lambda \alpha$ system
is quite large. When the Scheerbaum factor evaluated in the Born
approximation is used in the Scheerbaum potential,
one needs to multiply $S_\Lambda$ by
factor, $(\zeta+4)/4=1.297$, where $\zeta=M_\Lambda/M_N$ is
the mass ratio of $\Lambda$ to the nucleon.
A more precise comparison can therefore be made
by using $S^{\rm eff}_\Lambda=(1+\zeta/4)S_\Lambda
=-10.12~\hbox{MeV}\cdot \hbox{fm}^5$ from $S_\Lambda
=-7.8~\hbox{MeV}\cdot \hbox{fm}^5$. The latter value is
obtained from the $\bp=0$ Wigner transform
with $\bar{q}=0.7~\hbox{fm}^{-1}$ in Ref.\ \cite{SPLS}.
This potential is plotted in Fig.\ \ref{fig1} with
the dotted curve. If we use the Scheerbaum potential
with this $S^{\rm eff}_\Lambda$ value,
we obtain $\Delta E_{\ell s}=147$ keV, which is close to 137 keV.
Similarly, the fss2 value,
$S_\Lambda=-10.87~\hbox{MeV}\cdot \hbox{fm}^5$ or
$S^{\rm eff}_\Lambda=-14.10~\hbox{MeV}\cdot \hbox{fm}^5$,
in the Born approximation yields $\Delta E_{\ell s}=204$ keV.
It is amazing that the nonlocal $\Lambda \alpha$ kernel by FSS,
appearing in Fig.\ \ref{fig1} as the strongly
momentum-dependent Wigner transform, is well simulated
by a single-Gaussian Scheerbaum potential with
an appropriate $S^{\rm eff}_\Lambda$.
Figure \ref{fig1} also shows
the $\Lambda \alpha$ spin-orbit potential predicted
by the $G$-matrix simulated NSC97f-type $\Lambda N$ potential
in Ref.\ \cite{HI00}. The Scheerbaum factor
for this $\Lambda N$ potential is calculated
to be $S_\Lambda=-10.34~\hbox{MeV}\cdot \hbox{fm}^5$
for $\bar{q}=0.7~\hbox{fm}^{-1}$.
If we use the Scheerbaum potential
with $S^{\rm eff}_\Lambda=-13.41~\hbox{MeV}\cdot\hbox{fm}^5$,
we obtain $\Delta E_{\ell s}=194$ keV, which is close
to 209 keV.

Table \ref{table3} lists the results of $G$-matrix calculations
for the Scheerbaum factors $S_\Lambda$ in
symmetric nuclear matter.
The Fermi momentum, $k_F=1.07~\hbox{fm}^{-1}$, corresponding to
the half of the normal density $\rho_0=0.17~\hbox{fm}^{-3}$, is assumed.
For solving $G$-matrix equations, the continuous prescription
is used for intermediate spectra.
Table \ref{table3} also shows the decompositions
into various contributions and the results
when the $\Lambda N$-$\Sigma N$ coupling through
the $LS^{(-)}$ and $LS^{(-)}\sigma$ forces
is switched off (coupling off) in the $G$-matrix
calculations.
For FSS, we find a large reduction of $S_\Lambda$ value
from the Born value $-7.8~\hbox{MeV}\cdot \hbox{fm}^5$,
especially when this (dominantly) $P$-wave $\Lambda N$-$\Sigma N$ coupling
is properly taken into account.
When all the $\Lambda N$-$\Sigma N$ couplings, including those by
the pion tensor force, is switched off,
the $LS^{(-)}$ contribution is just a half
of the $LS$ contribution in the odd partial wave,
which is the same result as in the Born approximation.
The $P$-wave $\Lambda N$-$\Sigma N$ coupling enhances
the repulsive $LS^{(-)}$ contribution largely.
If we use this reduction of the $S_\Lambda$ factor
from $-7.8~\hbox{MeV}\cdot \hbox{fm}^5$
to $-1.9~\hbox{MeV}\cdot \hbox{fm}^5$ in the
realistic $G$-matrix calculation,
we find that the present $\Delta E_{\ell s}$ value $-137$ MeV is reduced
to an almost correct value, $-33$ keV.
However, such a reduction of the Scheerbaum factor due to
the $\Lambda N$-$\Sigma N$ coupling is
supposed to be hindered in the $\Lambda \alpha$ system
in the lowest-order approximation from the isospin consideration.
On the other hand, the situation of fss2 in Table \ref{table3} is
rather different, although the cancellation mechanism
between the $LS$ and $LS^{(-)}$ components and the reduction
effect of $S_\Lambda$ factor in the full calculation are
equally observed.
When all the $\Lambda N$-$\Sigma N$ coupling
is neglected, the ratio of the $LS^{(-)}$ and $LS$ contributions
in the quark sector is still one half.
Since the EMEP contribution is mainly for the $LS$ type,
it amounts to about $-6~\hbox{MeV}\cdot \hbox{fm}^5$,
which is very large and remains with the same magnitude
even after the $P$-wave $\Lambda N$-$\Sigma N$ coupling
is included. Furthermore, the increase of the $LS^{(-)}$ component
is rather moderate, in comparison with the FSS case.
This is becuse the model fss2 contains an appreciable EMEP
contribution ($\sim 40\%$) which has very
little $LS^{(-)}$ contributions.
As the result, the total $S_\Lambda$ value
in fss2 $G$-matrix calculation is 3 - 6 times
larger than the FSS value, depending on the Fermi
momentum, $k_F=1.35$ - $1.07~\hbox{fm}^{-1}$.
Such an appreciable EMEP contribution
to the $LS$ component of the $YN$ interaction is
not favorable to reproduce the negligibly small $\ell s$ splitting
of $\hbox{}^9_\Lambda \hbox{Be}$.

\begin{table}[t]
\caption{The Scheerbaum factors $S_\Lambda$ in
symmetric nuclear matter with $k_F=1.07~\hbox{fm}^{-1}$,
predicted by $G$-matrix calculations of FSS and fss2
in the continuous prescription for intermediate spectra.
Decompositions into various contributions are shown,
together with the cases
when the $\Lambda N$-$\Sigma N$ coupling by
the $LS^{(-)}$ and $LS^{(-)}\sigma$ forces is
switched off (coupling off).
The unit is in $\hbox{MeV}\cdot \hbox{fm}^5$.}
\label{table3}
\begin{center}
\renewcommand{\arraystretch}{1.4}
\setlength{\tabcolsep}{2mm}
\begin{tabular}{ccrrrr}
\hline
model & & \multicolumn{2}{c}{full}
 & \multicolumn{2}{c}{coupling off} \\
\hline
      & & odd & even & odd & even \\
\hline
     & $LS$       & $-17.36$ & 0.38 & $-19.70$ & 0.30 \\
FSS  & $LS^{(-)}$ &  14.83   & 0.22 &    8.37  & 0.26 \\
     & total & \multicolumn{2}{c}{$-1.93$}
     & \multicolumn{2}{c}{$-10.77$} \\
\hline
     & $LS$       & $-19.97$ & $-0.14$ & $-21.04$ & $-0.20$ \\
fss2 & $LS^{(-)}$ &    8.64  &    0.21 &     6.12 &    0.23 \\
     & total & \multicolumn{2}{c}{$-11.26$}
     & \multicolumn{2}{c}{$-14.89$} \\
\hline
\end{tabular}
\end{center}
\end{table}

Summarizing this work, we have performed the $jj$-coupling
Faddeev calculations for $\hbox{}^9_\Lambda \hbox{Be}$,
by incorporating $\Lambda \alpha$ $LS$ interactions
generated from the Born kernel of the quark-model
baryon-baryon interactions.
This calculation corresponds to the evaluation 
of the Scheerbaum factors in the Born approximation.
Since the $P$-wave $\Lambda N$-$\Sigma N$ coupling
is not properly taken into account,
the present calculation using the FSS Born kernel yields too large
spin-orbit splitting of the $5/2^+$ and $3/2^+$ excited states
of $\hbox{}^9_\Lambda \hbox{Be}$ by factor three.
In the model FSS, a reduction by factor 1/2 - 1/4 is
expected in the $G$-matrix calculation of the Scheerbaum
factor $S_\Lambda$ \cite{SPLS}, depending on
the Fermi momentum, $k_F=1.35$ - $1.07~\hbox{fm}^{-1}$.
In fss2, the $G$-matrix calculation for the Scheerbaum
factor yields a rather large value,
$S_\Lambda \sim -11~\hbox{MeV}\cdot \hbox{fm}^5$,
with very weak $k_F$ dependence, due to the appreciable
EMEP contributions. The quark-model baryon-baryon
interaction with a large spin-orbit contribution from the
meson-exchange potentials is, in general, unfavorable
to reproduce the very small $\ell s$ splitting observed
in $\hbox{}^9_\Lambda \hbox{Be}$.
It is a future problem how to incorporate
the $P$-wave $\Lambda N$-$\Sigma N$ coupling
in cluster model calculations like the present one.  


\begin{acknowledgments}
This work was supported by Grants-in-Aid for Scientific
Research (C) from the Japan Society for the Promotion
of Science (JSPS) (Nos.~15540270, 15540284, and 15540292).
\end{acknowledgments}


\begin{thebibliography}{9}
\bibitem{BE9L} Y. Fujiwara, K. Miyagawa, M. Kohno, Y. Suzuki,
D. Baye, and J.-M. Sparenberg,
Phys. Rev. C {\bf 70}, No. 2 (2004) (in press),
KUNS-1910, nucl-th/0404071.
\bibitem{fss2} Y. Fujiwara, T. Fujita, M. Kohno, C. Nakamoto,
and Y. Suzuki, Phys. Rev. C {\bf 65}, 014002 (2002).
\bibitem{B8B8} Y. Fujiwara, M. Kohno, C. Nakamoto, and Y. Suzuki,
Phys. Rev. C {\bf 64}, 054001 (2001).
\bibitem{SB97} J.-M. Sparenberg and D. Baye,
Phys. Rev. C {\bf 55}, 2175 (1997).
\bibitem{AK02} H. Akikawa {\em et al.}, Phys. Rev. Lett. {\bf 88},
082501 (2002).
\bibitem{TA03} H. Tamura {\em et al.}, in {\em Proceedings
of the VIII-th International Conference on Hypernuclear and
Strangeness Particle Physics} (HYPER2003),
Jefferson Lab, Newport News, Virginia, October 14-18, 2003,
to be published in Nucl. Phys. A (2004).
\bibitem{FSS} Y. Fujiwara, C. Nakamoto, and Y. Suzuki,
Phys. Rev. Lett. {\bf 76}, 2242 (1996);
Phys. Rev. C {\bf 54}, 2180 (1996).
\bibitem{NA93} C. Nakamoto, Y. Suzuki and Y. Fujiwara,
Phys. Lett. {\bf B318}, 587 (1993).
\bibitem{GMAT} M. Kohno, Y. Fujiwara, T. Fujita,
C. Nakamoto, and Y. Suzuki,
Nucl. Phys. {\bf A674}, 229 (2000).
\bibitem{SPLS} Y. Fujiwara, M. Kohno, T. Fujita,
C. Nakamoto, and Y. Suzuki,
Nucl. Phys. {\bf A674}, 493 (2000).
\bibitem{HI00} E. Hiyama, M. Kamimura, T. Motoba, T. Yamada,
and Y. Yamamoto, Phys. Rev. Lett. {\bf 85}, 270 (2000).
\bibitem{triton} Y. Fujiwara, K. Miyagawa, M. Kohno, Y. Suzuki,
and H. Nemura, Phys. Rev. C {\bf 66}, 021001(R) (2002).
\bibitem{hypt} Y. Fujiwara, K. Miyagawa, M. Kohno, and Y. Suzuki,
Phys. Rev. C\,{\bf 70},\,No.2\,(2004) (in press), nucl-th/0404010.
\bibitem{HE6LL} Y. Fujiwara, M. Kohno, K. Miyagawa, Y. Suzuki,
and J.-M. Sparenberg, KUNS-1920, nucl-th/0405056,
submitted to Phys. Rev. C.
\bibitem{LSRGM} Y. Fujiwara, M. Kohno, T. Fujita, C. Nakamoto,
and Y. Suzuki, Prog. Theor. Phys. {\bf 103}, 755 (2000).
\bibitem{HI97} E. Hiyama, M. Kamimura, T. Motoba, T. Yamada,
and Y. Yamamoto, Prog. Theor. Phys. {\bf 97}, 881 (1997).
%
\end{thebibliography}
\end{document}